\documentclass[aps,amsmath,amssymb,preprintnumbers,a4paper,prd,twocolumn]{revtex4-1}
 \pdfoutput=1

 \usepackage{lipsum}
\usepackage{mathrsfs}  

\usepackage{slashed}
\usepackage{subfigure}
\usepackage{feynmp}
\usepackage{graphicx}
\usepackage{amsfonts}
\usepackage{color}
\usepackage{cancel}

\def\hbar{\hspace{0pt}\raisebox{1pt}{$-$} \hspace{-7pt} h}

\def\5{\overline 5}

\definecolor{JJ}{RGB}{0,144,255}

\newcommand{\be}{\begin{equation}}
\newcommand{\ee}{\end{equation}}
\newcommand{\bea}{\begin{eqnarray}}
\newcommand{\eea}{\end{eqnarray}}

\newcommand{\ba}{\begin{eqnarray}}
\newcommand{\ea}{\end{eqnarray}}

\begin{document}
\title{Diboson Signals via Fermi Scale Spin-One States }

\author{Diogo Buarque Franzosi}

\author{Mads T. Frandsen}

\author{Francesco Sannino}
 \affiliation{CP$^{3}$-Origins and the Danish IAS, University of Southern Denmark, Campusvej 55, DK-5230 Odense M, Denmark}

\begin{abstract} 

ATLAS and CMS observe deviations from the expected background  in diboson invariant mass  searches of new resonances around 2 TeV. We provide a general analysis of the results in terms of spin-one resonances and find that Fermi scale composite dynamics can be the culprit. The analysis and methodology   can be employed for future searches at run two of the Large Hadron Collider.

\preprint{CP3-Origins-2015-023 DNRF90, DIAS-2015-23}

\end{abstract}

\maketitle

 The ATLAS search for diboson resonances using boson-tagged jets \cite{Aad:2015owa} finds local excesses of 3.4, 2.6 and 2.9 $\sigma$ in the $WZ$, $WW$ and $ZZ$ tagged boosted dijets with invariant mass spectrum around 2 TeV. This leads to a a global significance of 2.5$\sigma$. 
Similarly CMS finds an excess of 1.9 $\sigma$ global significance in a boosted search for $WH$ with the Higgs decaying hadronically \cite{CMS:2015gla}. 

The ATLAS results in the $VV'$ gauge boson channels suggest a reconstructed mass of around $2$~TeV  and a cross-section $\sigma (pp \to R\to VV')\equiv \sigma_ {VV'} $ of the order of $(6-10) ~{\rm fb} $. Here $R$ denotes a new intermediate massive vector boson and $V,V'$ weak gauge bosons. 
The lower value corresponds to the maximisation of the likelihood model based on Poisson statistics for the $W'$ peak described in the ATLAS reference. 
\footnote{We use the likelihood 
\be
L=\prod_i P_{poi}(n_{obs}^i|n_{exp}^i)
\ee
where $P_{poi}(n|\lambda)$ is the Poisson distribution of $n$ for a mean value $\lambda$, the index $i$ correspond to the bins in the reconstructed di-boson mass distribution of the $WZ$ channel, Fig. 5(a) in Ref.\cite{Aad:2015owa}, centred from $m_{jj}=1750$ GeV to $2350$ GeV, $n_{obs}$ are the number of events observed in the respective bin and $n_{exp}$ is the number of events expected for the sum of Standard Model processes plus the peak contribution from the $W'$ model used in the experimental paper scaled with a signal strength $\mu$. 
}
The upper value is extracted from the most stringent upper limits provided by other di-boson searches in ATLAS and CMS, which is provided by ATLAS semi-leptonic channel \cite{Aad:2015ufa} and is still in good agreement with the peak excess found by ATLAS.
$\sigma_ {VV'}=(6-10)$ fb will be our {\it region of interest}.

 Here we employ a minimal description of spin-one resonances and study their phenomenology in the narrow width approximation. The model encompasses, however, all the needed ingredients to describe the {\it signal channels} and relevant constraints.
We then make contact with time-honoured models of minimal composite dynamics \cite{Weinberg:1975gm,Susskind:1978ms,Kaplan:1983fs,Kaplan:1983sm}. Weinberg and Susskind's minimal models of weak scale, also known as the Fermi scale, composite dynamics \cite{Weinberg:1975gm,Susskind:1978ms}  are based on QCD-like dynamics and  are at odds with experiments. On the other hand, modern incarnations that are still  minimal but employ non-QCD like dynamics are phenomenologically viable \cite{Sannino:2004qp,Dietrich:2005jn,Ryttov:2008xe}.  The associated {\it signal channels} have  been investigated in more complete model implementations, e.g. in \cite{Belyaev:2008yj}.

Although the data are not yet conclusive, the general features, regarding resonance mass, cross-section and decay patterns are very much in line with models of weak scale compositeness \cite{Belyaev:2008yj}.

 \vskip .2cm
{\it Spin-one Lagrangian:}
Since we are interested in the hadronic production and diboson decays of spin-one resonances $R$ we consider the simplified effective Lagrangian 
\bea 
{\cal L}^{R}={\cal L}^{R}_{qq}+{\cal L}^{R}_{VV}+{\cal L}^{R}_{VH}+{\cal L}^{R}_{\mathcal{X}}
\eea
where $qq$ denotes quarks, and $H$ is the observed Higgs state while $\mathcal{X}$ is everything else, e.g leptons and dark matter. 
Correspondingly the width of the resonance can be written as 
\bea
\Gamma_R=\Gamma_{qq}+\Gamma_{VV}+ \Gamma_{VH} + \Gamma_\mathcal{X} 
 \eea

The vertices linking the spin-one resonances with the standard model fermions are 
\bea
\mathcal{L}^{R}_{\rm qq}&=\sum_{u,d} \bar{u}\slashed{{R}}^+\left(g^V_{ u d}- g^A_{ u d}\, \gamma_5\right) d + {\rm h.c.}
\\
&+\sum_{ij}  \bar{q}_i \slashed{{R}}^0 \left(g^V_{ij }-g^A_{ij}\, \gamma_5\right) q_j \ ,
\eea
where $u$ ($d$) runs over all up-type (down-type) quarks and $q$ runs over all quark flavors, and we also further make the simplifying assumption $g^{V/A}_{ij} =g^{V/A}_{ud}  =g_{{V/A}}$. 
 
Neglecting $CP$-violating terms (see e.g.~\cite{Hagiwara:1986vm} for a
more complete discussion) the couplings of neutral $R\equiv R_\mu^0$ to standard model gauge fields can
be written as
\bea
\mathcal{L}^{R}_{VV} &=& g^{R}_{WW1} [[RW^+W^-]]_1 + g^{R}_{WW2} [[RW^+W^-]]_2 
\nonumber \\ 
& + &  g^{R}_{WW3} [[RW^+W^-]]_\epsilon + g^{R}_{ZZ} [[R ZZ]]_\epsilon
\label{eq:gauge}
\eea
where
\bea
[[RW^+W^-]]_1 & \equiv & 2 i\left[ \partial_{[\mu} W_{\nu]}^+  W^{\mu -} R^{\nu} - \partial_{[\mu} W_{\nu]}^- W^{\mu +} R^{\nu}\right] \;, \nonumber \\{}
[[RW^+W^-]]_2 & \equiv & \frac{i}{2}(\partial_\mu R_\nu - \partial_\nu R_\mu) (W^{\mu +} W^{\nu -} - W^{\mu -} W^{\nu +})  \;,\nonumber\\{}
[[RV_1V_2]]_\epsilon & \equiv & \epsilon^{\mu\nu\rho\sigma}   (V_{1 \mu} \partial_\rho V_{2 \nu}- \partial_\rho V_{1\mu}  V_{2\nu})  R_\sigma \;, \nonumber \\
\eea
We also have  ${L}^{R}_{ZH} = g^{R}_{ZH}  R_{\mu}  Z^{\mu} H$.   An equivalent Lagrangian can be defined for the charged spin-one resonances which we omit for brevity. 

\begin{figure}[htb!]  
\begin{center}
 \includegraphics[width=.7\columnwidth]{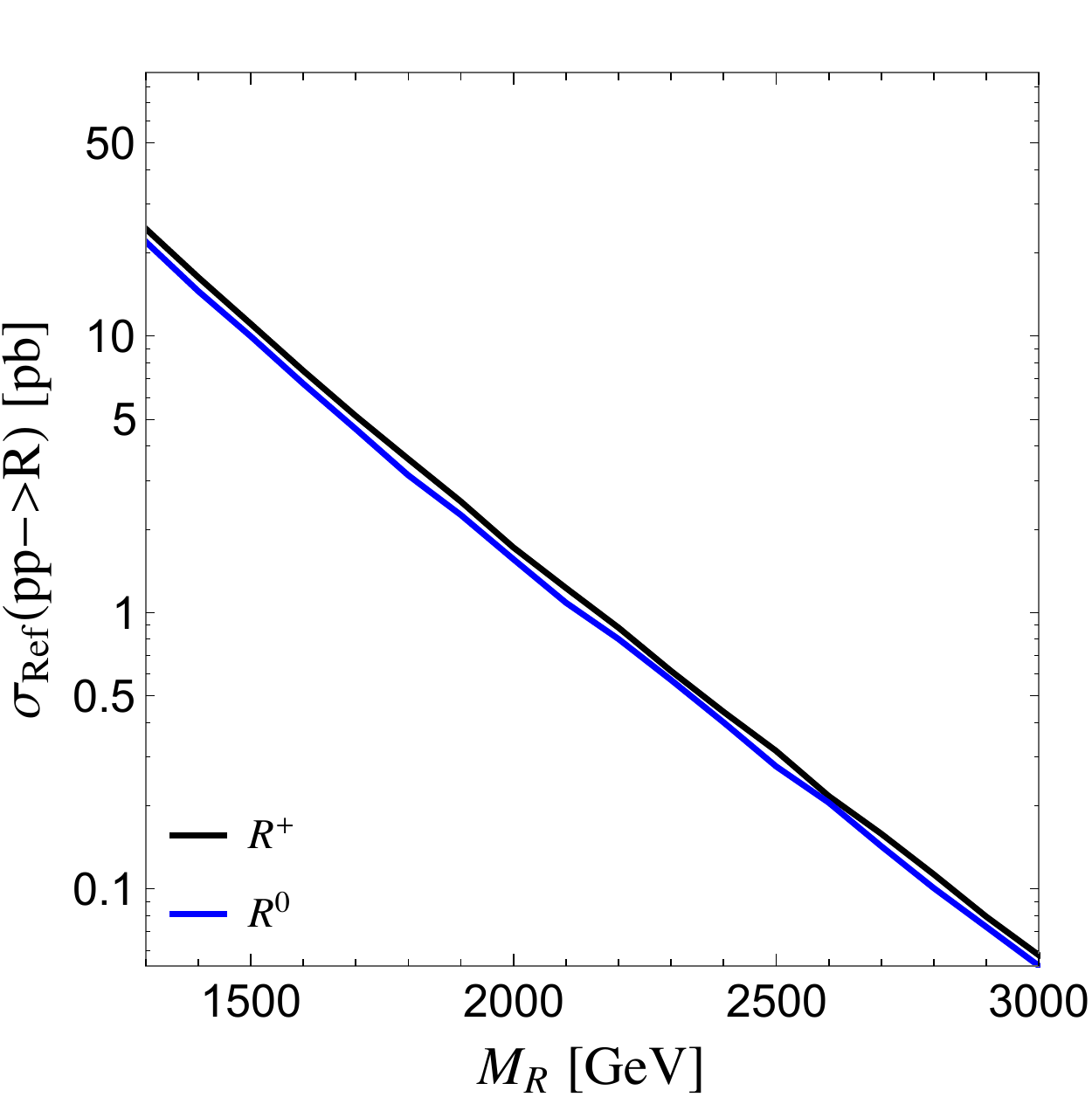}
 \caption{ Production cross-section $ \sigma_{\rm Ref} (pp\to R) $ with $g_V = 1$  and $g_A = 0$ for the couplings to fermions.   }
\label{fig:cs}
\end{center}
\end{figure}

 \vskip .2cm
{\it Production cross section and widths:}
We now show in Fig.~\ref{fig:cs} the  reference production cross-sections for a neutral and charged spin-one resonance 
 with $g_V = 1$  and $g_A=0$. We used the NNPDF2.3 set of parton distribution functions \cite{Ball:2012cx} and the {\sc MadGraph5\_aMC@NLO} framework \cite{Alwall:2014hca} to get cross sections, with the Universal Feynrules Model \cite{Degrande:2011ua}  described in \cite{Chiang:2011kq}.
For a 2 TeV vector mass we have   \begin{eqnarray}
\sigma_{\rm Ref}(pp\to R^{0}) & = & 1.5 \times 10^3 fb \ , \quad m_R^ 0 = 2~{\rm TeV}\nonumber \\
\sigma_{\rm Ref} (pp\to R^{\pm}) & = & 1.7 \times 10^3 fb  \ , \quad m_R^\pm = 2~{\rm TeV}
\label{fiducial}
\end{eqnarray}
The partial widths  for the neutral vector are well approximated by 
\begin{eqnarray}
\Gamma(R\rightarrow q\bar{q})&\simeq &\frac{m_R N_c}{12\pi} 
[(g^{V})^2+(g^{A})^2]   \ , \nonumber \\ 
 \Gamma(R\rightarrow W^{+}W^{-}) & \simeq &
\frac{1}{192  \pi}m_R \left (\frac{m_R}{m_W} \right )^4
(g_{WW2}^R)^2   \ , \nonumber \\
 \Gamma(R\rightarrow ZW) & \simeq &
\frac{1}{192  \pi}m_R \left (\frac{m_R}{m_W} \right )^4
(g_{ZW2}^R)^2   \ , \nonumber \\
 \Gamma(R\rightarrow Z Z ) &\simeq& 
\frac{(g_{ZZ}^R)^2}{96  \pi}m_R \frac{m_R^2}{m_Z^2}  \ , \nonumber \\ 
\Gamma(R\rightarrow ZH)& \simeq & \frac{(g^R_{ZH})^2}{192\pi m_Z^2}  m_R \ .
\end{eqnarray}
 Here $N_c =3$ is the number of colors and in the following we will set $g_A =0$.

  \vskip .2cm
{\it  Constraints:}
 Given the production cross-sections \eqref{fiducial}, the total cross-section into diboson final states is
\bea
\sigma_{VV} &=& g_V^2 \times Br[R \to VV] \times \sigma_{\rm Ref}(pp\to R) \nonumber
\\  & = &c_{VV} \sigma_{\rm Ref}(pp\to R) ,
\eea 
where e.g. $c_{WZ} \sim (3-6)\times 10^{-3}$, in order to have $\sigma_{WZ}\sim (6-10)$ fb. In particular  we must have $g_V^2 \gtrsim c_{WZ} $ once we require a certain diboson cross-section.

 We now determine the lower limit on the resonance width, required to explain the excesses using both dijets  and dibosons (into boosted jets) searches.  The dijet  cross-section is
\bea
\sigma_{qq}  & =& g_V^2 \times Br[R\to qq] \times \sigma_{\rm Ref}(pp\to R) \nonumber
\\ & =& c_{qq}\sigma_{\rm Ref}(pp\to R) \ .
\eea 
for which the current limits at 2 TeV require $c_{qq}\lesssim 0.1$  \cite{Khachatryan:2015sja}. It follows that the total width is 
\bea
\frac{\Gamma_{R_V}}{m_{R_V}} &\gtrsim \frac{\Gamma_{qq}}{m_{R_V}} (1 +  \frac{c_{WZ}+c_{WH}}{c_{qq}}) + \frac{\Gamma_{\mathcal{X}}}{m_{R_V}} \nonumber 
\\
&\simeq g_V^2 \frac{N_f \times 3}{12 \pi }(1 +  \frac{c_{WZ}+c_{WH}}{c_{qq}})+ \frac{\Gamma_\mathcal{X}}{m_R} ,
\eea 
where we have taken $N_f$ light quark flavors. So again requiring $\sigma_{WZ}\sim (6-10)$ fb we find $\frac{\Gamma_{R_V}}{m_{R_V}} \gtrsim (8-15) \times 10^{-4}$. 
Taking into account the dilepton decay modes, included in $\Gamma_\mathcal{X}$, will add a subdominant contribution to the width.  The resulting limit shows that the narrow width approximation is justified.

\begin{figure*}[htb!] 
\begin{center}
  \includegraphics[width=.4\textwidth]{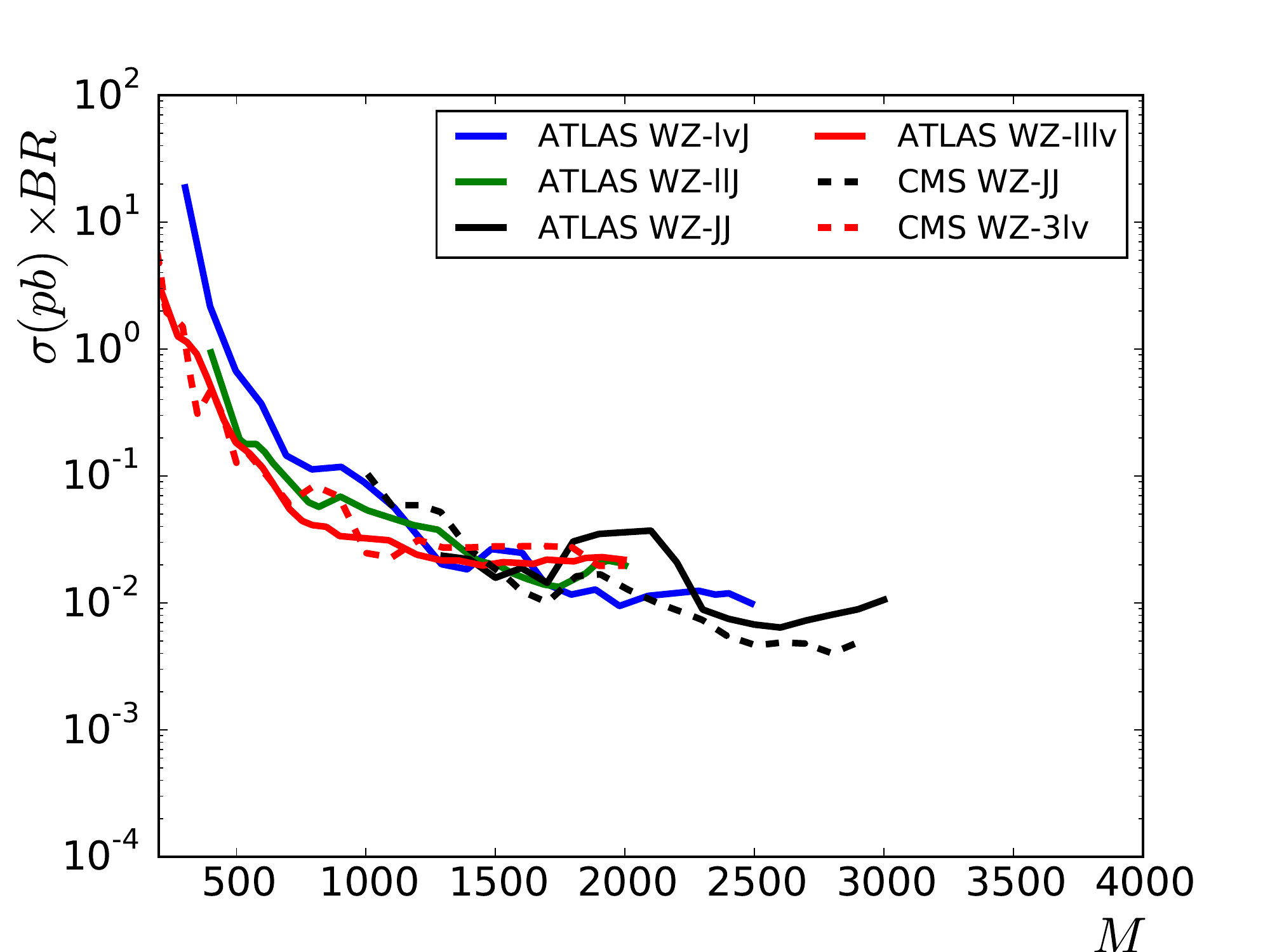}
 \includegraphics[width=.4\textwidth]{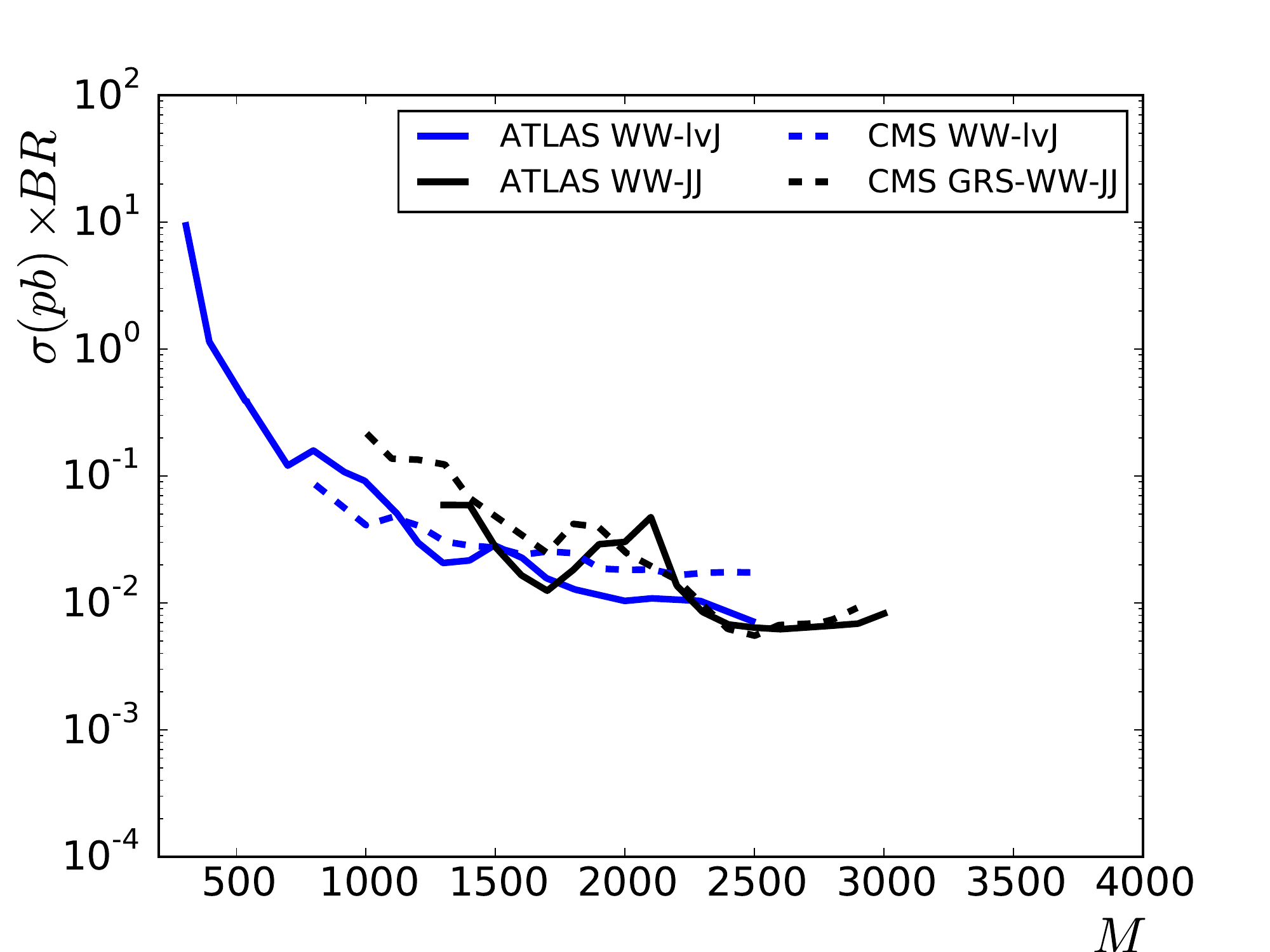}
 \includegraphics[width=.4\textwidth]{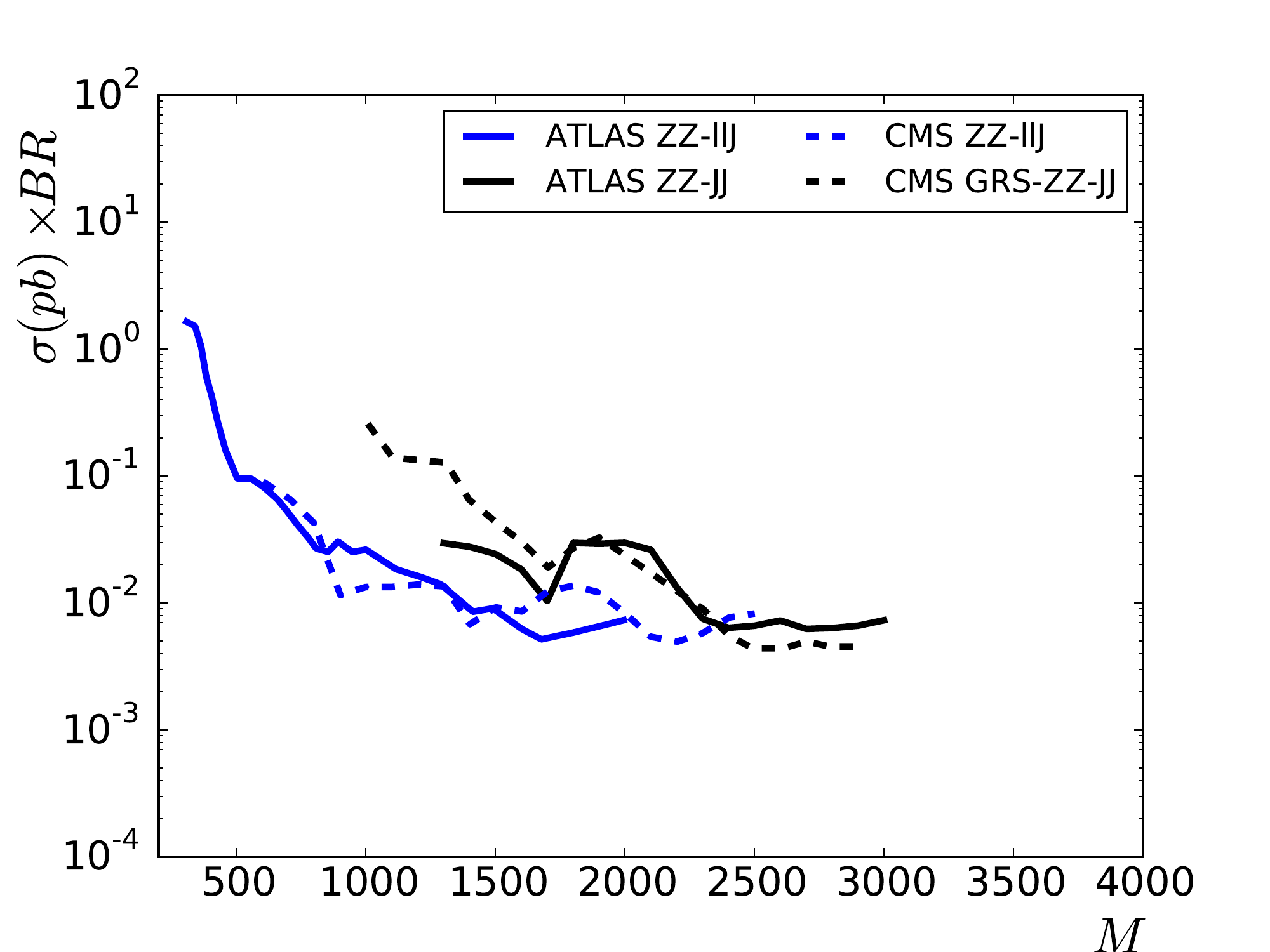}
 \includegraphics[width=.4\textwidth]{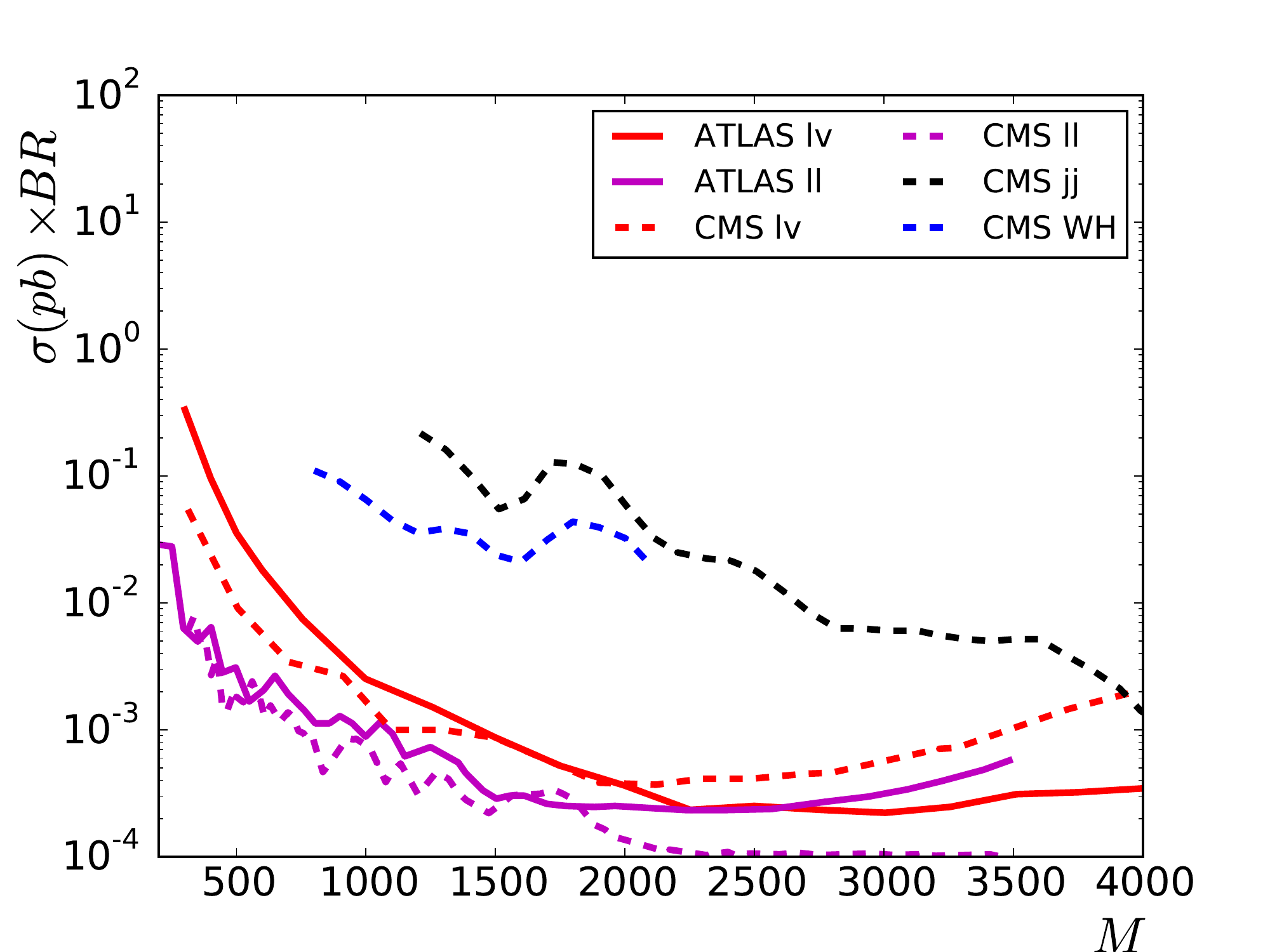}
\caption{ATLAS and CMS 95\% CL exclusion limits on the production cross-section times branching ratio, $\sigma \times BR$, for a new neutral or charged vector resonance. A capital $J$ indicates  $W$ or  $Z$ reconstructed via jets. \emph{Upper left:} $WZ$ searches. \emph{Upper right:} $WW$ searches. \emph{Bottom left:} $ZZ$ searches. \emph{Bottom right:} di-lepton, di-jets and $WH$ searches. The experimental references are \cite{Khachatryan:2015sja,Khachatryan:2014fba,Khachatryan:2014tva,CMS:2015gla,Khachatryan:2014hpa,Khachatryan:2014gha,Khachatryan:2014xja,Aad:2015owa,Aad:2015ufa,Aad:2014xka,Aad:2014pha}.  }
\label{fig:xsbr}
\end{center}
\end{figure*}

A second constraint arises from dilepton searchers.  Of course this is a model dependent constraint that is, for example,  absent in leptophobic models.    The current LHC limits for a single charged or neutral vector resonance are given in Fig.~\ref{fig:xsbr}. We refer to \cite{Becciolini:2014eba} for a discussion of the details of the plot.

We conclude from the plot that the current limits on $\sigma(pp\to R\to \ell \nu)$ at $m_R\simeq 2$ TeV translate into the constraint $g_V^2 \times Br[R\to \ell \nu] \lesssim 5 \times 10^{-4}$. Using that $g_V^2 \gtrsim c_{WZ} \sim (3-6)\times 10^{-3}$ we have that  $Br[R\to \ell \nu] \lesssim (16-8) \times 10^{-2}$. The conclusions are similar for the neutral resonances, with respect to dijet and dilepton final states.  Later we shall see that such a constraint is naturally abided by minimal models of weak composite dynamics. 

Other leptonic and semileptonic searches are weaker than the ATLAS semi-leptonic limit on $\sigma_{WZ}$. 
The ATLAS fully leptonic search for resonances decaying into $WZ$ \cite{Aad:2014pha}  limits the cross-section $\sigma(pp\to R\to WZ)\lesssim 22$ fb at $m_R=2$ TeV. However it does not extend beyond 2 TeV.  
ATLAS semi-leptonic and CMS fully hadronically boosted analyses for resonances decaying into $WZ$ give upper limits for the total cross section of around 10fb and 12fb respectively. Similarly in $WW$ and $ZZ$ semi-leptonic searches the upper limit on the total cross sections are higher than the semi-leptonic limits. 

\vskip .2cm
{\it  $ZZ$ - challenge: }
Given that the spin-one states are weak triplets it is natural to expect signals in  $WZ$ and $WW$ channels. On the other hand, large contributions to  the $ZZ$ decay mode requires anomalous couplings violating $P$ and/or $CP$. Given that the mass resolution in the jet mass reconstruction of the $W$ and $Z$ is  $\pm 13$~GeV, it is logically possible that the ZZ reconstructed events do in fact involve $W$s. In the future with sufficient statistics in leptonic final states, this channel will be a diagnostic of the resonance nature. This  possibility is demonstrated in \cite{Fukano:2015hga}. 

\vskip .2cm
{\it  Fermi (Goldstone) composite dynamics and Lattice results: }
We now argue that models of  spin-one resonances from composite dynamics at the electroweak scale can have the required mass, production cross-section and partial widths to explain the observed excesses at ATLAS. 

For the spectrum we  use first principle lattice results  \cite{Lewis:2011zb, Fodor:2012ty,Hietanen:2013fya,Hietanen:2014xca}.  Specifically in models based on $SU(3)$ with fermions in the 2-index symmetric representation \cite{Sannino:2004qp} recent lattice results indicate that the lightest vector and axial triplets have masses of \cite{Fodor:2012ty}
$m_{R_V}\simeq 1.75 \pm 0.1$ TeV, $m_{R_A}\simeq 2.3 \pm 0.1$ TeV.  

Lattice results for $SU(2)$ driven composite dynamics with fermions in the fundamental representation \cite{Appelquist:1999dq, Duan:2000dy,Ryttov:2008xe}  yield \cite{Hietanen:2014xca}
$m_{R_V}\simeq 2.5 \pm 0.5$ TeV, $m_{R_A}\simeq 3.3 \pm 0.7$ TeV. 

It is worth mentioning that even the simple scaled up version of QCD suggests the existence of a spin-one vector of mass around  2~TeV. 

The examples above assume the electroweak condensate scale of $246$~GeV to be identified with the decay constant of the composite pions, i.e. the ones absorbed in the longitudinal degrees of freedom of the massive weak gauge bosons. For the $SU(2)$ fundamental dynamics case with chiral symmetry breaking pattern $SU(4) \to Sp(4)$,  however, one can imagine a more general electroweak embedding, parametrised by an angle $ 0\leq \theta \leq \pi/2$,  that for $\theta$ close to zero yields the composite Goldstone Higgs scenario \cite{Kaplan:1983fs,Kaplan:1983sm}. Recent realisations were considered in \cite{Katz:2005au,Gripaios:2009pe,Galloway:2010bp,Barnard:2013zea,Ferretti:2013kya,Cacciapaglia:2014uja}.  As function of the embedding angle we have $m_R(\theta) = m_R/\sin \theta$  with $m_R$ given above  for $\theta = \pi/2$ \cite{Cacciapaglia:2014uja}.  If, for example, one requires some typical values of $\theta \simeq 10^{-1}$, the lattice predicted vector resonances states would be too heavy to be observable at the LHC.  A comprehensive phenomenological analysis of the intriguing scalar sector of the theory \cite{Cacciapaglia:2014uja} appeared in \cite{Arbey:2015exa} for any value of theta.

Other underlying theories are also being  studied on the lattice, see for example \cite{Lucini:2015noa}. 

Further, in minimal weak scale composite models spin-one resonances couple to standard model fermions via mixing  with the electroweak  gauge bosons. Additional model building can also yield direct couplings to the fermions. From the mixing alone one deduces $g_V \sim \frac{g^2}{\sqrt{2}\tilde{g}}$ 
where one power of $g$ comes from the $W$ gauge eigenstate coupling to $ud$ while $\frac{g}{\tilde{g}}\sim  \frac{m_W}{m_R}$ arises from the mass mixing. $\tilde{g}$ is the self-coupling of the new spin-one mesons.  Further ${\cal O}(1)$ corrections depend on the parameters of the mass mixing Lagrangian \cite{Belyaev:2008yj}.  Using the value of the weak coupling we have $g_V\sim {0.25}/{\tilde{g}}$ and therefore to achieve $g_V^2 Br[R\to WZ]=c_{WZ}\sim 3-6 \times 10^{-3}$  implies $\tilde{g}\lesssim 4.5$. This is a very natural value of $\tilde{g}$ since for a composite spin-one state we expect  $1\lesssim \tilde{g}\lesssim 4\pi$. In QCD, for example its value is about $2\pi$. 

Early studies of weak scale minimal composite dynamics \cite{Belyaev:2008yj}  find the spin-one resonance width $\Gamma_R\sim 10^2$ GeV at 2 TeV for $\tilde{g} \sim 5$ with dilepton branching ratios at the level of $10^{-3}$such that $g_V^2 Br[R\to \ell \ell] \sim 10^{-5}-10^{-4}$  \cite{Belyaev:2008yj}.

Moreover we expect both vector and axial spin-one triplets with  (mainly) axial weak triplets having further significant decay modes into $HW$ and $HZ$ final states \cite{Belyaev:2008yj}. 

{Flavor constraints have been discussed in much detail in \cite{Fukano:2009zm} for theories featuring spin-one resonances. }

Amusingly the lattice results for the spin-one spectrum encompas the one needed to explain the experimental excesses and the deduced couplings to standard model fermions are naturally of the expected order of magnitude. 

The run two experiment at the LHC  will be able to either confirm or dismiss this intriguing possibility.

 \vspace{.2cm}

{\it Acknowledgments:} We thank Georges Azuelos, Angel Campoverde, Flavia Dias and Robert McCarthy for  information and clarifications about the experimental results. The CP$^3$-Origins center is partially funded by the Danish National Research Foundation, grant number DNRF90. 

\vspace{.2cm}
{\it Note Added:} While this work was being finalised the analysis \cite{Fukano:2015hga} appeared.  Our discussion of spin-one resonances is fairly  general and does not rely on a specific underlying model. Moreover  we use lattice data that provide the spectrum of spin-one resonances relevant to experiments.


\begin{thebibliography}{999}                                                                                               

\bibitem{Aad:2015owa} 
  G.~Aad {\it et al.}  [ATLAS Collaboration],
  arXiv:1506.00962 [hep-ex].


\bibitem{CMS:2015gla} 
  CMS Collaboration [CMS Collaboration],
  CMS-PAS-EXO-14-010.


\bibitem{Weinberg:1975gm} 
  S.~Weinberg,
  Phys.\ Rev.\ D {\bf 13}, 974 (1976).


\bibitem{Susskind:1978ms} 
  L.~Susskind,
  Phys.\ Rev.\ D {\bf 20}, 2619 (1979).


\bibitem{Kaplan:1983fs} 
  D.~B.~Kaplan and H.~Georgi,
  Phys.\ Lett.\ B {\bf 136}, 183 (1984).


\bibitem{Kaplan:1983sm} 
  D.~B.~Kaplan, H.~Georgi and S.~Dimopoulos,
  Phys.\ Lett.\ B {\bf 136}, 187 (1984).


\bibitem{Sannino:2004qp} 
  F.~Sannino and K.~Tuominen,
  Phys.\ Rev.\ D {\bf 71}, 051901 (2005)
  [hep-ph/0405209].


\bibitem{Dietrich:2005jn} 
  D.~D.~Dietrich, F.~Sannino and K.~Tuominen,
  Phys.\ Rev.\ D {\bf 72}, 055001 (2005)
  [hep-ph/0505059].


\bibitem{Ryttov:2008xe} 
  T.~A.~Ryttov and F.~Sannino,
  Phys.\ Rev.\ D {\bf 78}, 115010 (2008)
  [arXiv:0809.0713 [hep-ph]].


\bibitem{Belyaev:2008yj} 
  A.~Belyaev, R.~Foadi, M.~T.~Frandsen, M.~Jarvinen, F.~Sannino and A.~Pukhov,
  Phys.\ Rev.\ D {\bf 79}, 035006 (2009)
  [arXiv:0809.0793 [hep-ph]].


\bibitem{Hagiwara:1986vm} 
  K.~Hagiwara, R.~D.~Peccei, D.~Zeppenfeld and K.~Hikasa,
  Nucl.\ Phys.\ B {\bf 282}, 253 (1987).


\bibitem{Ball:2012cx} 
  R.~D.~Ball, V.~Bertone, S.~Carrazza, C.~S.~Deans, L.~Del Debbio, S.~Forte, A.~Guffanti and N.~P.~Hartland {\it et al.},
  Nucl.\ Phys.\ B {\bf 867}, 244 (2013)
  [arXiv:1207.1303 [hep-ph]].


\bibitem{Alwall:2014hca} 
  J.~Alwall, R.~Frederix, S.~Frixione, V.~Hirschi, F.~Maltoni, O.~Mattelaer, H.-S.~Shao and T.~Stelzer {\it et al.},
  JHEP {\bf 1407}, 079 (2014)
  [arXiv:1405.0301 [hep-ph]].


\bibitem{Degrande:2011ua} 
  C.~Degrande, C.~Duhr, B.~Fuks, D.~Grellscheid, O.~Mattelaer and T.~Reiter,
  Comput.\ Phys.\ Commun.\  {\bf 183}, 1201 (2012)
  [arXiv:1108.2040 [hep-ph]].


\bibitem{Chiang:2011kq} 
  C.~W.~Chiang, N.~D.~Christensen, G.~J.~Ding and T.~Han,
  Phys.\ Rev.\ D {\bf 85}, 015023 (2012)
  [arXiv:1107.5830 [hep-ph]].


\bibitem{Khachatryan:2015sja} 
  V.~Khachatryan {\it et al.}  [CMS Collaboration],
  Phys.\ Rev.\ D {\bf 91}, no. 5, 052009 (2015)
  [arXiv:1501.04198 [hep-ex]].


\bibitem{Becciolini:2014eba} 
  D.~Becciolini, D.~B.~Franzosi, R.~Foadi, M.~T.~Frandsen, T.~Hapola and F.~Sannino,
  arXiv:1410.6492 [hep-ph].


\bibitem{Khachatryan:2014fba} 
  V.~Khachatryan {\it et al.}  [CMS Collaboration],
  JHEP {\bf 1504}, 025 (2015)
  [arXiv:1412.6302 [hep-ex]].


\bibitem{Khachatryan:2014tva} 
  V.~Khachatryan {\it et al.}  [CMS Collaboration],
  Phys.\ Rev.\ D {\bf 91}, no. 9, 092005 (2015)
  [arXiv:1408.2745 [hep-ex]].


\bibitem{Khachatryan:2014hpa} 
  V.~Khachatryan {\it et al.}  [CMS Collaboration],
  JHEP {\bf 1408}, 173 (2014)
  [arXiv:1405.1994 [hep-ex]].


\bibitem{Khachatryan:2014gha} 
  V.~Khachatryan {\it et al.}  [CMS Collaboration],
  JHEP {\bf 1408}, 174 (2014)
  [arXiv:1405.3447 [hep-ex]].


\bibitem{Khachatryan:2014xja} 
  V.~Khachatryan {\it et al.}  [CMS Collaboration],
  Phys.\ Lett.\ B {\bf 740}, 83 (2015)
  [arXiv:1407.3476 [hep-ex]].


\bibitem{Aad:2015ufa} 
  G.~Aad {\it et al.}  [ATLAS Collaboration],
  Eur.\ Phys.\ J.\ C {\bf 75}, no. 5, 209 (2015)
  [arXiv:1503.04677 [hep-ex]].


\bibitem{Aad:2014xka} 
  G.~Aad {\it et al.}  [ATLAS Collaboration],
  Eur.\ Phys.\ J.\ C {\bf 75}, no. 2, 69 (2015)
  [arXiv:1409.6190 [hep-ex]].


\bibitem{Aad:2014pha} 
  G.~Aad {\it et al.}  [ATLAS Collaboration],
  Phys.\ Lett.\ B {\bf 737}, 223 (2014)
  [arXiv:1406.4456 [hep-ex]].


\bibitem{Fukano:2015hga} 
  H.~S.~Fukano, M.~Kurachi, S.~Matsuzaki, K.~Terashi and K.~Yamawaki,
  arXiv:1506.03751 [hep-ph].


\bibitem{Lewis:2011zb} 
  R.~Lewis, C.~Pica and F.~Sannino,
  Phys.\ Rev.\ D {\bf 85}, 014504 (2012)
  [arXiv:1109.3513 [hep-ph]].


\bibitem{Fodor:2012ty} 
  Z.~Fodor, K.~Holland, J.~Kuti, D.~Nogradi, C.~Schroeder and C.~H.~Wong,
  Phys.\ Lett.\ B {\bf 718}, 657 (2012)
  [arXiv:1209.0391 [hep-lat]].


\bibitem{Hietanen:2013fya} 
  A.~Hietanen, R.~Lewis, C.~Pica and F.~Sannino,
  JHEP {\bf 1412}, 130 (2014)
  [arXiv:1308.4130 [hep-ph]].


\bibitem{Hietanen:2014xca} 
  A.~Hietanen, R.~Lewis, C.~Pica and F.~Sannino,
  JHEP {\bf 1407}, 116 (2014)
  [arXiv:1404.2794 [hep-lat]].


\bibitem{Appelquist:1999dq} 
  T.~Appelquist, P.~S.~Rodrigues da Silva and F.~Sannino,
  Phys.\ Rev.\ D {\bf 60}, 116007 (1999)
  [hep-ph/9906555].


\bibitem{Duan:2000dy} 
  Z.~y.~Duan, P.~S.~Rodrigues da Silva and F.~Sannino,
  Nucl.\ Phys.\ B {\bf 592}, 371 (2001)
  [hep-ph/0001303].


\bibitem{Katz:2005au} 
  E.~Katz, A.~E.~Nelson and D.~G.~E.~Walker,
  JHEP {\bf 0508}, 074 (2005)
  [hep-ph/0504252].


\bibitem{Gripaios:2009pe} 
  B.~Gripaios, A.~Pomarol, F.~Riva and J.~Serra,
  JHEP {\bf 0904}, 070 (2009)
  [arXiv:0902.1483 [hep-ph]].


\bibitem{Galloway:2010bp} 
  J.~Galloway, J.~A.~Evans, M.~A.~Luty and R.~A.~Tacchi,
  JHEP {\bf 1010}, 086 (2010)
  [arXiv:1001.1361 [hep-ph]].


\bibitem{Barnard:2013zea} 
  J.~Barnard, T.~Gherghetta and T.~S.~Ray,
  JHEP {\bf 1402}, 002 (2014)
  [arXiv:1311.6562 [hep-ph]].


\bibitem{Ferretti:2013kya} 
  G.~Ferretti and D.~Karateev,
  JHEP {\bf 1403}, 077 (2014)
  [arXiv:1312.5330 [hep-ph]].


\bibitem{Cacciapaglia:2014uja} 
  G.~Cacciapaglia and F.~Sannino,
  JHEP {\bf 1404}, 111 (2014)
  [arXiv:1402.0233 [hep-ph]].


\bibitem{Arbey:2015exa} 
  A.~Arbey, G.~Cacciapaglia, H.~Cai, A.~Deandrea, S.~Le Corre and F.~Sannino,
  arXiv:1502.04718 [hep-ph].


\bibitem{Lucini:2015noa} 
  B.~Lucini,
  arXiv:1503.00371 [hep-lat].

\bibitem{Fukano:2009zm} 
  H.~S.~Fukano and F.~Sannino,
  Int.\ J.\ Mod.\ Phys.\ A {\bf 25}, 3911 (2010)
  [arXiv:0908.2424 [hep-ph]].

\end{thebibliography}
\end{document}